\documentclass[useAMS,usenatbib,usegraphicx]{mn2e}
\title[Direct reconstruction of $a_{\pm2,lm}$ with interferometers]
{Direct reconstruction of spherical harmonics from interferometer observations
of the CMB polarization}
\author[Jaiseung Kim]{Jaiseung Kim$^{1}$\thanks{E-mail:
jkim@physics.brown.edu}\\ 
$^{1}$Dept. of Physics, Brown Univ. Providence, U.S.A.}

\begin{document}
\date{Accepted 2006 November 12. Received 2006 November 2; in original form 2006 September 10}
\pagerange{\pageref{firstpage}--\pageref{lastpage}} \pubyear{}
\maketitle
\label{firstpage}
\begin{abstract}
Interferometric observation of the CMB polarization can be expressed as a linear sum of 
spherical harmonic coefficients $a_{\pm 2,lm}$ of the CMB polarization. The linear weight 
for $a_{\pm 2,lm}$ depends on the observational configuration such as antenna pointing, baseline orientation, and spherical harmonic number $l,m$.
Since an interferometer is sensitive over a finite range of multipoles, $a_{\pm 2,lm}$ in the range can be determined by fitting $a_{\pm 2,lm}$
for visibilities of various observational configurations. 
The formalism presented in this paper enables the determination of $a_{\pm 2,lm}$ directly from spherical harmonic spaces without spherical harmonic transformation of pixellized maps.
The result of its application to a simulated observation is presented with the formalism.
\end{abstract}

\begin{keywords}
-- cosmology: cosmic microwave background -- techniques: interferometric -- methods: data analysis
\end{keywords}
 
\section{Introduction}
The Cosmic Microwave Background (CMB) is expected to be linearly polarized by Thomson scattering at the last scattering surface and after re-ionization.
The CMB polarization has been measured by the \textit{DASI} \citep{DASI:3yr}, the \textit{CBI} \citep{CBI:polarization}, the \textit{BOOMERanG} \citep{BOOMERANG:Polarization}, the \textit{CAPMAP} \citep{CAPMAP:polarization} and the \textit{WMAP} satellite \citep{WMAP:polarization}.
A characteristic signature imprinted  on CMB polarization provides valuable cosmological and astrophysical information.
If the CMB anisotropy follows Gaussian distribution, the complete description of the CMB anisotropy is provided through the angular power spectrum $C_{l}$ \citep{Modern_Cosmology,WMAP:3yr_TT}. 

Interferometers offer more control of systematic effects than traditional imaging systems, and have less E and B mode mixing \citep{Park:EB_separation}.
With desirable features of interferometers, the CMB polarization measurement with interferometers is on-going and planned in the experiments such as the \textit{DASI} \citep{DASI:data,DASI:instrument,DASI:3yr}, the \textit{Cosmic Background Imager} (\textit{CBI}) \citep{CBI:polarization} and the \textit{Millimeter-wave Bolometric Interferometer} (\textit{MBI}) \citep{Tucker:Bolometric_interferometry,SPIE:MBI}.
The usual procedure for the CMB analysis on the interferometer observation is to proceed
to a statistical analysis such as maximum likelihood estimation of power spectra.
In the power spectrum estimation by maximum likelihood method, 
$\mathcal O(\mathcal N^3)$ process should be carried out repeatedly for iterative search.
Since it becomes computationally prohibitive with very large number of data, an unbiased hybrid estimator \citep{hybrid_estimation} proposes pseudo-$C_l$ estimates at high multipoles, which requires the estimation of individual spherical harmonic coefficients. 
Spherical harmonic coefficients can be estimated also from mosaiced 
sky patches, which are reconstructed from interferometer observations via aperture synthesis. Due to flat sky approximation for each sky patch, the mosaiced sky map has discontinuity on junctures of sky patches. 
For these reasons, we have investigated reconstructing the spherical harmonic coefficients of the CMB polarization directly from interferometer observations in the complete context of a spherical sky. 

This paper is organized as follows.
We discuss Stokes parameters in \S 2. Interferometric CMB polarization measurement on spherical sky is discussed in \S 3. 
In \S 4, we show visibilities are linearly weighted sum of spin $\pm$2 spherical harmonic coefficients. In \S 5, we show how spin $\pm$2 spherical harmonic coefficients can be determined from visibilities. In \S 6, computational feasibility is discussed. 
In \S 7, reconstruction results from simulated observations are presented.
In \S 8, the summary and discussion are given. 
In Appendix A, we discuss methods to facilitate computation of linear weight for $a_{\pm2,lm}$.
In Appendix B, the reconstruction results without noise is presented. 
\section{ALL-SKY STOKES PARAMETERS}
\label{Stokes}
There are Stokes parameter Q and U, which describe the state of polarization \citep{Kraus:Radio_Astronomy,Tools_Radio_Astronomy}. Since Thomson scattering does not generate circular polarization in early Universe, circular polarization state $V$ is not considered here.
In this paper, we follow the polarization convention of the 
\begin{scriptsize}HEALPIX\end{scriptsize}                                          \citep{HEALPix:framework},  which differs from the definition of the International Astronomical Union.
In all-sky analysis, these are measured in reference to $(\mathbf{\hat e_\theta},\mathbf{\hat e_\phi})$ \citep{Zaldarriaga:Polarization_Exp,Seljak-Zaldarriaga:Polarization}. $\mathbf{\hat e_\theta}$ and $\mathbf{\hat e_\phi}$ are unit vectors of a spherical coordinate system and given by \citep{Arfken}
\begin{eqnarray*}
\mathbf {\hat e_{\theta}}&=&\mathbf {\hat i}\;\cos\theta\cos\phi+\mathbf {\hat j}\;\cos\theta\sin\phi-\mathbf {\hat k}\;\sin\theta,\\
\mathbf {\hat e_{\phi}}&=&-\mathbf {\hat i}\;\sin\phi+\mathbf {\hat j}\;\cos\phi.
\end{eqnarray*}

Stokes parameters Q and U are as follows:
\begin{eqnarray} 
\label{Q} Q&=&\left\langle E_\theta^2-E_\phi^2\right\rangle,\\
\label{U} U&=&\left\langle 2E_\theta\,E_\phi\right\rangle,
\end{eqnarray}
where $\left\langle \ldots\right\rangle$ indicates time average.
$Q$ and $U$ transform under rotation of an angle $\psi$ on the plane perpendicular to direction $\mathbf {\hat n}$ as  
\begin{eqnarray}
\label{Q'} Q'(\mathbf {\hat n})&=&Q(\mathbf {\hat n})\cos2\psi+U(\mathbf {\hat n})\sin2\psi,\\
\label{U'} U'(\mathbf {\hat n})&=&-Q(\mathbf {\hat n})\sin2\psi+U(\mathbf {\hat n})\cos2\psi,
\end{eqnarray}
with which the following quantities can be constructed \citep{Seljak-Zaldarriaga:Polarization,Zaldarriaga:Polarization_Exp}:
\begin{eqnarray}
\label{Q'U'} (Q\pm \mathrm{i} U)'(\mathbf {\hat n})=\mathrm e^{\mp 2\mathrm{i}\psi}(Q\pm \mathrm{i} U)(\mathbf {\hat n}).
\end{eqnarray}

For all-sky analysis, $Q$ and $U$ are expanded in terms of spin $\pm2$ spherical harmonics \citep{Seljak-Zaldarriaga:Polarization,Zaldarriaga:Polarization_Exp,Zaldarriaga:thesis} as follows:
\begin{eqnarray}
\label{Q_lm+iU_lm} Q(\hat {\mathbf n})+\mathrm{i} U(\hat {\mathbf n})&=&\sum_{l,m} a_{2,lm}\;{}_2Y_{lm}(\hat {\mathbf n}),\\
\label{Q_lm-iU_lm} Q(\hat {\mathbf n})-\mathrm{i} U(\hat {\mathbf n})&=&\sum_{l,m} a_{-2,lm}\;{}_{-2}Y_{lm}(\hat {\mathbf n}),
\end{eqnarray}
$a_{2,lm}$ is related to $a_{-2,lm}$  by $a_{-2,lm}=(-1)^m a_{2,l\,-m}^*$ \citep{Seljak-Zaldarriaga:Polarization}. 
Spin $\pm2$ spherical harmonics have following forms:
\begin{eqnarray}
{}_2Y_{lm}(\theta,\phi)&=&\sqrt{\frac{2l+1}{4\pi}}[F_{1,lm}(\theta)+F_{2,lm}(\theta)]e^{\mathrm{i} m\phi},\label{2Ylm}\\
{}_{-2}Y_{lm}(\theta,\phi)&=&\sqrt{\frac{2l+1}{4\pi}}[F_{1,lm}(\theta)-F_{2,lm}(\theta)]e^{ \mathrm{i} m\phi},\label{-2Ylm}
\end{eqnarray}
where $F_{1,lm}$ and $F_{2,lm}$ can be computed in terms of Legendre functions as follows \citep{Kamionkowski:Flm,Zaldarriaga:Polarization_Exp}:
\begin{eqnarray}
F_{1,lm}(\theta)&=&2\sqrt{\frac{(l-2)!(l-m)!}{(l+2)!(l+m)!}}[(l+m)\frac{\cos\theta}{\sin^2\theta}P^m_{l-1}(\cos\theta)\nonumber\\
&&-(\frac{l-m^2}{\sin^2\theta}+\frac{1}{2}l(l-1))P^m_l(\cos\theta)],\label{F1lm}\\
F_{2,lm}(\theta)&=&2\sqrt{\frac{(l-2)!(l-m)!}{(l+2)!(l+m)!}}\frac{m}{\sin^2\theta}[(l+m)P^m_{l-1}(\cos\theta)\nonumber\\
&&-(l-1)\cos\theta P^m_l(\cos\theta)].\label{F2lm}
\end{eqnarray}
\section{Interferometric Measurement}
The discussion in this section is for an ideal interferometer.
An interferometer measures time-averaged correlation of two electric field from a pair of identical apertures positioned at $\mathbf r_1$ and at $\mathbf r_2$. 
The separation, $\mathbf B=\mathbf r_1-\mathbf r_2$, of two apertures is called the `baseline' and the measured correlation is called `visibility' \citep{Lawson:Interferometry,Thompson:interferometer}.
Depending on the instrumental configuration, visibilities
are associated with $\langle E_x^2-E_y^2\rangle$, $\langle 2 E_x E_y\rangle$ and $\langle E_x^2-E_y^2\pm \mathrm{i}\,2 E_x E_y\rangle$ respectively, where $\hat x$ and $\hat y$ are axes of the polarizer frame.
As discussed in \S \ref{Stokes}, Stokes parameters at angular coordinate ($\theta$,$\phi$) are defined in respect to two basis vectors $\mathbf {\hat e_\theta}$ and $\mathbf {\hat e_\phi}$.
Consider the polarization observation, whose antenna pointing is in the direction of angular coordinate ($\theta_A$,$\phi_A$). The polarizers and baselines are assumed to be on the aperture plane. Then, the global frame coincides with the polarizer frame after Euler rotation ($\phi_A$, $\theta_A$, $\psi$) on the global frame, where $\psi$ is the rotation around the axis in the direction of antenna pointing.
Most of interferometer experiments for the CMB observation employ feedhorns for beam collection.
After passing through the feedhorn system, an incoming off-axis ray becomes on-axis ray.
Then, the basis vectors $\mathbf {\hat e_\theta}$ and $\mathbf {\hat e_\phi}$ of the ray after the feedhorn system are related to the basis vectors $\mathbf {\hat e_x}$ and $\mathbf {\hat e_y}$ 
of the polarizer frame as follows:
\begin{eqnarray}
\label{azimuthal_rotation}
\mathbf {\hat e}_{x}+\mathrm{i}\,\mathbf {\hat e}_{y}=
\mathrm e^{-\mathrm{i}\psi}(\mathbf {\hat e}_{\theta_A}+\mathrm{i}\,\mathbf {\hat e}_{\phi_A})
=\mathrm e^{\mathrm{i}(\Phi-\psi)}(\mathbf {\hat e}_\theta+\mathrm{i}\,\mathbf {\hat e}_\phi),
\end{eqnarray}
where $\Phi$ is given by 
\begin{eqnarray*}
\label{Phi}\Phi&=&\tan^{-1}\left[\frac{\sin\theta\sin(\phi-\phi_A)}{\sin\theta\cos\theta_A\cos(\phi-\phi_A)-\cos\theta\sin\theta_A}\right]\\
&&+\tan^{-1}\left[\frac{\sin\theta_A\sin(\phi-\phi_A)}{-\sin\theta\cos\theta_A+\cos\theta\sin\theta_A\cos(\phi-\phi_A)} \right].
\end{eqnarray*}
Refer to Appendix \ref{relation} for the details on the derivation of $\Phi$.
With Eq. \ref{azimuthal_rotation}, we can easily show that
\begin{eqnarray*}
\langle E_x^2- E_y^2\rangle+\mathrm{i}\langle 2E_x\,E_y\rangle=e^{-\mathrm{i}(2\psi-2\Phi)}
(\langle E_\theta^2- E_\phi^2\rangle+\mathrm{i}\langle 2E_\theta\,E_\phi\rangle).
\end{eqnarray*}
With the employment of linear polarizers, the visibilities are associated with $\langle E_x^2-E_y^2\rangle$ or $\langle 2 E_x E_y\rangle$, and are as follows:
\begin{eqnarray}
\label{V_Q_spherical}
V_{Q^\prime}&=&f(\nu)\int \mathrm d \Omega A(\mathbf {\hat n},\hat {\mathbf n}_A)\\
&&\times\mathrm{Re}\,[\mathrm e^{-\mathrm{i}(2\psi-2\Phi(\mathbf {\hat n}))}
(Q(\mathbf {\hat n})+\mathrm{i} U(\mathbf {\hat n}))]\,\mathrm e^{\mathrm{i}\,2\pi\mathbf u\cdot \mathbf {\hat n}},\nonumber\\
\label{V_U_spherical}
V_{U^\prime}&=&f(\nu)\int \mathrm d \Omega A(\mathbf {\hat n},\hat {\mathbf n}_A)\\
&&\times\mathrm{Im}\,[\mathrm e^{-\mathrm{i}(2\psi-2\Phi(\mathbf {\hat n}))}
(Q(\mathbf {\hat n})+\mathrm{i} U(\mathbf {\hat n}))]\,\mathrm e^{\mathrm{i}\,2\pi\mathbf u\cdot \mathbf {\hat n}},\nonumber
\end{eqnarray}
where $\hat {\mathbf n}_A$ is the  unit vector in the direction of antenna pointing and
$f(\nu)$ is the frequency spectrum of the CMB polarization. \footnote{$f(\nu)=\left.\frac {\partial B(\nu,T)} {\partial T}\right|_{T=T_0}$, where $B(\nu,T)$ is the Plank function and
$T_0$ is the CMB monopole temperature.}
With the employment of circular polarizers, the visibilities are associated with $\langle E_x^2-E_y^2\pm \mathrm{i}\,2 E_x E_y\rangle$, and are as follows: 
\begin{eqnarray}
\label{V_RL_spherical}
V_{RL}&=&f(\nu)\int \mathrm d \Omega A(\mathbf {\hat n},\hat {\mathbf n}_A),\\ 
&&\times [Q(\mathbf {\hat n})+\mathrm{i} U(\mathbf {\hat n})]\mathrm e^{\mathrm{i}(2\pi\mathbf u\cdot \mathbf {\hat n}-2\psi+2\Phi(\mathbf {\hat n}))},\nonumber\\
\label{V_LR_spherical}
V_{LR}&=&f(\nu)\int \mathrm d \Omega A(\mathbf {\hat n},\hat {\mathbf n}_A),\\
&&\times [Q(\mathbf {\hat n})-\mathrm{i} U(\mathbf {\hat n})]\mathrm e^{\mathrm{i}(2\pi\mathbf u\cdot \mathbf {\hat n}+2\psi-2\Phi(\mathbf {\hat n}))},\nonumber
\end{eqnarray}
where $R$ and $L$ stand for right/left circular polarizers.

As in the following, $V_{Q^\prime}$ and $V_{U^\prime}$ are linear combinations of
$V_{RL}$ and $V_{LR}$, and vice versa.
\begin{eqnarray}
V_{Q^\prime}&=&\frac{1}{2}(V_{RL}+V_{LR}),\label{V_Q_symmetry}\\
V_{U^\prime}&=&-\frac{\mathrm{i}}{2}(V_{RL}-V_{LR}),\label{V_U_symmetry}\\
V_{RL}&=&V_{Q^\prime}+\mathrm{i} V_{U^\prime},\label{V_RL_symmetry}\\
V_{LR}&=&V_{Q^\prime}-\mathrm{i} V_{U^\prime},\label{V_LR_symmetry}
\end{eqnarray}

\section{Visibility as the linear sum of spherical harmonic coefficients}
With Eq. \ref{Q_lm+iU_lm} and \ref{Q_lm-iU_lm}, CMB visibilities can be expressed as a linearly weighted sum of $a_{\pm 2,lm}$
in following ways:
\begin{eqnarray}
\label{V_Q_linear}
V_{Q^\prime}&=&\frac{1}{2}\sum_{l,m}(a_{2,lm}b_{2,lm}+a_{-2,lm}b_{-2,lm}),\\
\label{V_U_linear}
V_{U^\prime}&=&-\frac{\mathrm{i}}{2}\sum_{l,m}(a_{2,lm}b_{2,lm}-a_{-2,lm}b_{-2,lm}),\\
\label{V_RL_linear}
V_{RL}&=&\sum_{l,m}a_{2,lm}b_{2,lm},\\
\label{V_LR_linear}
V_{LR}&=&\sum_{l,m}a_{-2,lm}b_{-2,lm},
\end{eqnarray}
where
\begin{eqnarray}
b_{2,lm}&=&\int d \nu f(\nu)\int\mathrm d \Omega A(\hat {\mathbf n},\hat {\mathbf n}_A)\label{b_2lm}\\
&&\times \mathrm e^{\mathrm{i}(2\pi\mathbf u_i\cdot \hat {\mathbf n}-2\psi+2\Phi(\hat {\mathbf n}))}{}_2Y_{lm}(\hat {\mathbf n})\nonumber,\\
b_{-2,lm}&=&\int d \nu f(\nu)\int\mathrm d \Omega A(\hat {\mathbf n},\hat {\mathbf n}_A)\label{b_-2lm}\\
&&\times \mathrm e^{\mathrm{i}(2\pi\mathbf u_i\cdot \hat {\mathbf n}+2\psi-2\Phi(\hat {\mathbf n}))}{}_{-2}Y_{lm}(\hat {\mathbf n})\nonumber.
\end{eqnarray}
All the instrumental and configurational information are contained in $b_{\pm2,lm}$.
As seen in Eq. \ref{V_Q_linear}, \ref{V_U_linear}, \ref{V_RL_linear} and \ref{V_LR_linear},
$b_{\pm 2,lm}$ are linear weights for $a_{\pm 2,lm}$.
As seen in Eq.\ref{b_2lm} and \ref{b_-2lm}, $b_{\pm 2,lm}$ have distinct values, which depend on its spherical harmonic number $l,m$ and the observational configuration such as antenna pointing and baseline. 
\section{determination of individual spherical harmonic coefficient}
\label{determination}
An interferometer is sensitive to multipoles of a range $l_0\le l\le l_1$. 
$\frac{l_0+l_1}{2}$ is given by $2\pi u$, where $u$ is a baselinelength divided by a wavelength.
The width of the range, $l_1-l_0$, depends on the window function, which is 
the square of the beam function in spherical harmonic space.
When the interferometer is sensitive to multipole of a range $l_0\leq l\leq l_1$,
there are $(l_1+1)^2-l_0^2$ spin $\pm2$ spherical harmonics in the range. 
Spin $\pm2$ spherical harmonic coefficient $a_{\pm 2,lm}$ ($l_0\leq l\leq l_1$) can be determined by fitting them for visibilities of various antenna pointings and baseline orientations.
For coding convenience, we split visibilities, $b_{\pm 2,lm}$, and $a_{\pm 2,lm}$ into real and imaginary parts. We enumerate real and imaginary parts of visibilities, $b_{\pm 2,lm}$, and $a_{\pm 2,lm}$ in matrix notation as follows:
\begin{eqnarray}
\mathbf V=\mathbf b\,\mathbf a. \label{Vba}
\end{eqnarray}
Likelihood function is given by
\begin{eqnarray*}
\mathcal{L}&=&\frac{1}{(2\pi)^{n}(\det \mathbf N)^{\frac{1}{2}}}\exp[-\frac{1}{2}(\Delta-\mathbf b\,\mathbf a)^{\mathrm T}\mathbf N^{-1}(\Delta-\mathbf b\,\mathbf a)],
\end{eqnarray*}
where $n$ is the number of visibilities, $\Delta$ is a visibility data vector, $\mathbf N$ is a noise covariance matrix. 
The likelihood function is maximum at 
\begin{eqnarray}
\mathbf a=(\mathbf b^T \mathbf N^{-1} \mathbf b)^{-1}\mathbf b^T \mathbf N^{-1}\Delta. \label{maximum_likelihood}
\end{eqnarray}
Eq. \ref{maximum_likelihood} is reduced to $\mathbf a=\mathbf b^{-1}\Delta$
if b is square and b is invertible. 
The covariance of estimation error is
\begin{eqnarray}
\lefteqn{\langle \Delta\mathbf a\;\Delta \mathbf a^T\rangle}\nonumber\\
&=&\langle(\mathbf b^T \mathbf N^{-1} \mathbf b)^{-1}\mathbf b^T \mathbf N^{-1}\Delta_{N}\;\{(\mathbf b^T \mathbf N^{-1} \mathbf b)^{-1}\mathbf b^T \mathbf N^{-1}\Delta_{N}\}^T\rangle,\nonumber\\
&=&(\mathbf b^T\mathbf N^{-1} \mathbf b)^{-1},\label{error_covariance}
\end{eqnarray}
where $\Delta_{N}$ is the noise of a visibility data vector and $\langle \Delta_{N} {\Delta_{N}}^T\rangle=\mathbf N$.
The E and B decomposition modes can be determined as follows:
\begin{eqnarray}
a_{E,lm}&=&-(a_{2,lm}+(-1)^m a_{2,l\,-m}^*)/2\label{E_decomposition}\\
&=&-((-1)^m a_{-2,l-m}^*+ a_{-2,lm})/2,\nonumber
\end{eqnarray}
\begin{eqnarray}
a_{B,lm}&=&\mathrm i(a_{2,lm}-(-1)^m a_{2,l\,-m}^*)/2\label{B_decomposition}\\
&=&\mathrm i((-1)^m a_{-2,l-m}^*- a_{-2,lm})/2.\nonumber
\end{eqnarray}
The variance of $a_{E,lm}$ and $a_{B,lm}$ estimation error 
are
\begin{eqnarray}
\lefteqn{\langle \Delta a_{E,lm}\Delta a^*_{E,lm}\rangle}\label{E_variance}\\
&=&\frac{1}{4}(\langle \Delta a_{\pm2,lm}\Delta a^*_{\pm2,lm}\rangle+\langle \Delta a_{\pm2,l\,-m} \Delta a_{\pm2,l\,-m}^*\rangle\nonumber\\
&&+(-1)^m 2\,\mathrm {Re}[\langle \Delta a_{\pm2,l\,m} \Delta a_{\pm2,l\,-m}^*\rangle]
),\nonumber
\end{eqnarray}
\begin{eqnarray}
\lefteqn{\langle \Delta a_{B,lm}\Delta a^*_{B,lm}\rangle}\label{B_variance}\\
&=&\frac{1}{4}(\langle \Delta a_{\pm2,lm}\Delta a^*_{\pm2,lm}\rangle+\langle \Delta a_{\pm2,l\,-m} \Delta a_{\pm2,l\,-m}^*\rangle\nonumber\\
&&-(-1)^m 2\,\mathrm {Re}[\langle \Delta a_{\pm2,l\,m} \Delta a_{\pm2,l\,-m}^*\rangle]
),\nonumber
\end{eqnarray}
where the variance and covariance of $\Delta a_{\pm 2,l\,\pm m}$ are given by Eq. \ref{error_covariance}.
\section{scaling of computational load}
As shown in previous section, $a_{2,lm}$ ($l_0\leq l\leq l_1$) is determined by
\begin{eqnarray}
\mathbf a=(\mathbf b^T \mathbf N^{-1} \mathbf b)^{-1}\mathbf b^T \mathbf N^{-1}\Delta \label{a}.
\end{eqnarray}
$\mathbf a$ is the vector of length $\mathsf m$, $\mathbf b$ is a $\mathsf n\times \mathsf m$ matrix and $\mathbf N$ is a $\mathsf n\times \mathsf n$ matrix, where $\mathsf n$ is the number of visibilities and $\mathsf m$ is the number of $a_{2,lm}$, which is $(l_1+1)^2-l_0^2$.
Unlikes the instrumental noise of a single dish experiment, the noise covariance matrix for interferometric observations can be assumed to 
be diagonal \citep{CMB:strategy}.
Since inverting matrix is $\mathcal O(\mathcal N^3)$ \citep{Numerical_Recipe_C} while inverting diagonal matrix is $O(\mathcal N)$ operation, 
Eq. \ref{a} is a process of $\mathcal O(\mathsf m^3)$.
Computing $b_{\pm2,lm}$ is small computational load, compared with computing Eq. \ref{a}. The method to compute $b_{\pm2,lm}$ fast
is presented in Appendix \ref{computing_b2lm}.
Rough estimate by scaling our simulation in \S \ref{simulation} to higher multipole ($l\sim1500$) says it will take roughly $\sim 2500$ days by Pentium 4 (2Ghz) system.
With ultra performance of computers such as SGI or IBM, determination of 
$a_{2,lm}$ over high multipoles by this formalism is computationally feasible.

\section{simulated observation}
\label{simulation}
We used the CAMB \citep{CAMB} to compute the power spectra of $\Lambda CDM$ and tensor-to-scalar ratio ($r=0.3$). 
$a_{E,lm}$ and $a_{B,lm}$ sets are drawn from the CAMB power spectra.
With these $a_{E,lm}$ and $a_{B,lm}$, we have generated the simulated CMB Q and U maps by \[Q(\hat {\mathbf n})\pm i U(\hat {\mathbf n})=\sum_{l,m} a_{\pm2,lm}\;{}_{\pm2}Y_{lm}(\hat {\mathbf n}).\] 
With the Q and U maps, $V_{RL}$ and $V_{LR}$ were simulated by numerically computing the following:
\begin{eqnarray*}
V_{RL}&=&\mathrm{noise}+\int d \nu f(\nu)\int \mathrm d \Omega A(\mathbf {\hat n},\hat {\mathbf n}_A)\\ 
&&\times [Q(\mathbf {\hat n})+i U(\mathbf {\hat n})]\mathrm e^{i(2\pi\mathbf u_i\cdot \mathbf {\hat n}-2\psi+2\Phi(\mathbf {\hat n}))},\\
V_{LR}&=&\mathrm{noise}+\int d \nu f(\nu)\int \mathrm d \Omega A(\mathbf {\hat n},\hat {\mathbf n}_A)\\ 
&&\times [Q(\mathbf {\hat n})-i U(\mathbf {\hat n})]\mathrm e^{i(2\pi\mathbf u_i\cdot \mathbf {\hat n}+2\psi-2\Phi(\mathbf {\hat n}))}.
\end{eqnarray*}
We assumed the sensitivity of the \textit{Planck} at 30GHz \citep{Planck:sensitivity}: 
$13\,\mathrm{mJy}$ ($7.64\,\mu\mathrm{K}$) 
for noise, though nature of instruments are different. 
The observational frequency was assumed to be 30 -- 31 GHz with $30^\circ$ FWHM Gaussian primary beam \footnote{The conclusion of this paper is not affected by the shape of the beam function
as far as the window function corresponding to the beam function is not non-zero over infinite number of multipoles.}.
Total visibilities is $n_{\theta_A}\times  n_{\phi_A}\times n_{\phi_{\mathbf{u}}}$,
where the number of fields is $n_{\theta_A}\times  n_{\phi_A}$.
For each field, baseline orientations are assumed to be 
$\pi/n_{\phi_{\mathbf{u}}}\,k$, where $k=1,2,\cdots n_{\phi_{\mathbf{u}}}$.
This can be achieved by building a feedhorn array of a $n_{\phi_{\mathbf{u}}}$ fold rotational symmetry. The fields of survey are assumed to have angular coordinate $(\pi\;i/n_{\theta_A},2\pi \;j/n_{\phi_A})$, where $i=1,2,\cdots n_{\theta_A}$ and $j=1,2,\cdots n_{\phi_A}$.
As seen in Eq. \ref{b_2lm_baseline} and \ref{b_-2lm_baseline}, $b_{\pm2,lm}$ depends on antenna pointing ($\theta_A,\phi_A$) and baseline orientation $\phi_{\mathbf{u}}$. 
In simulated observation, we assumed $n_{\theta_A}= n_{\phi_A}= n_{\phi_{\mathbf{u}}}= n^{1/3}>((l_1+1)^2-l_0^2)^{1/3}$ for the variation of $(\theta_A,\phi_A,\phi_{\mathbf{u}})$.
\begin{figure}
\centering
\centering\includegraphics[scale=.5]{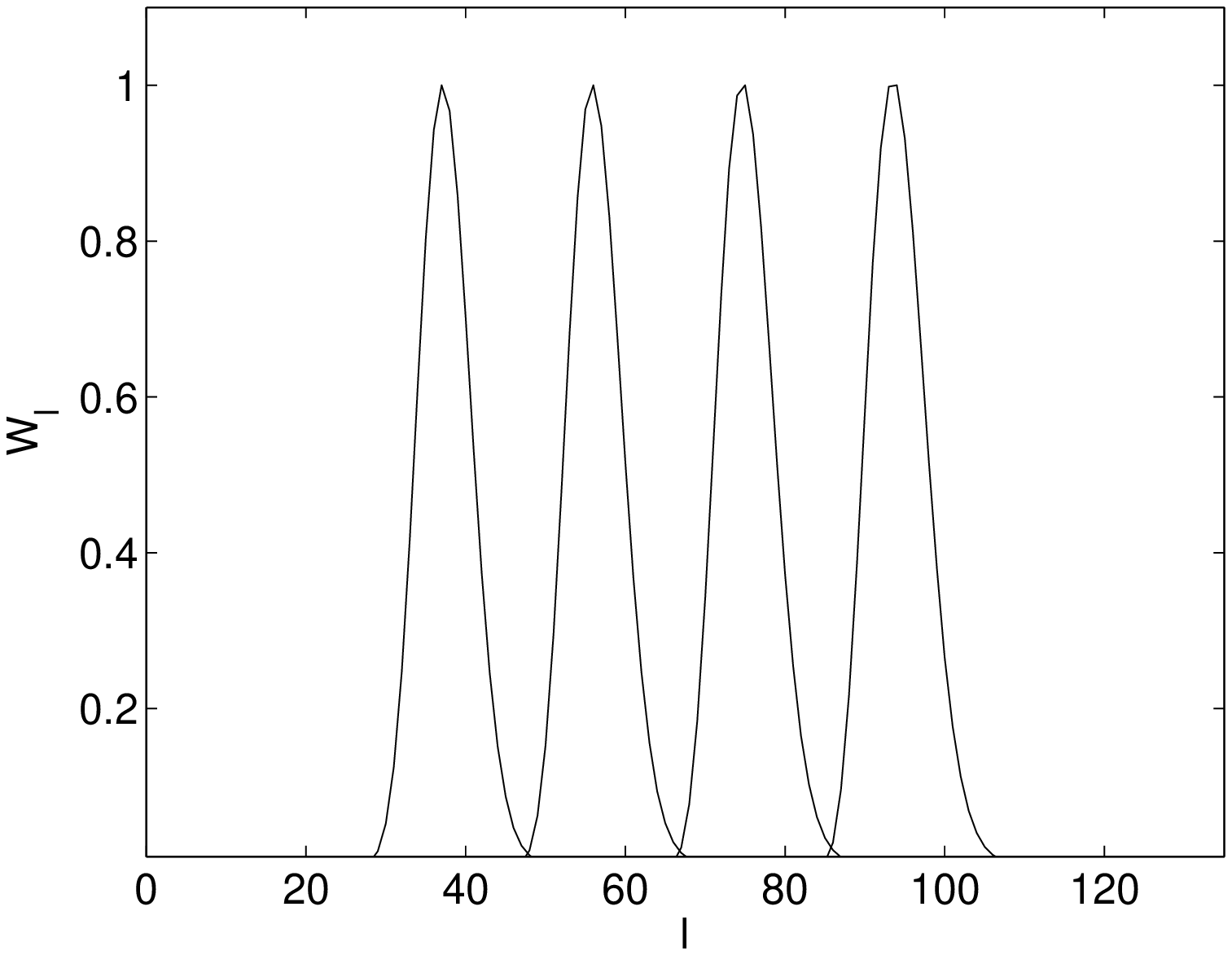}
\caption{Window Function (normalized to its peak.)}
\label{window_function}
\end{figure}

Baselines of length $B_1=6$ [cm], $B_2=9$ [cm], $B_3=12$ [cm] and $B_4=15$ [cm] with 
$30^\circ$ FWHM beams were assumed.
The corresponding window functions are shown in Fig. \ref{window_function}. 
A window function corresponding to the longest baseline is shown at rightmost.
The interferometers of $B_1$ are sensitive to the multipole range $29 \le l \le 48$, those of $B_2$ to $48 \le l \le 67$, those of $B_3$ to $67 \le l \le 86$ 
and those of $B_4$ to $86 \le l \le 106$.
We chose the multipole range $l_0$ and  $l_1$ to be the first multipoles where the window function drops below 1\% of its peak value. 
With such cutoff, there exists error from residuals, which contributes to estimation error.
We chose $n_{\theta_A}=n_{\phi_A}=n_{\phi_{\mathbf{u}}}=20$, which makes the total number of visibilities ($V_{RL}$ and $V_{LR}$) about four times the number of the spherical harmonics to be determined. So the number of constraints is about four times the number of unknowns, which is necessary in the presence of noise and residual error.

\begin{figure}
\centering
\centering\includegraphics[scale=.5]{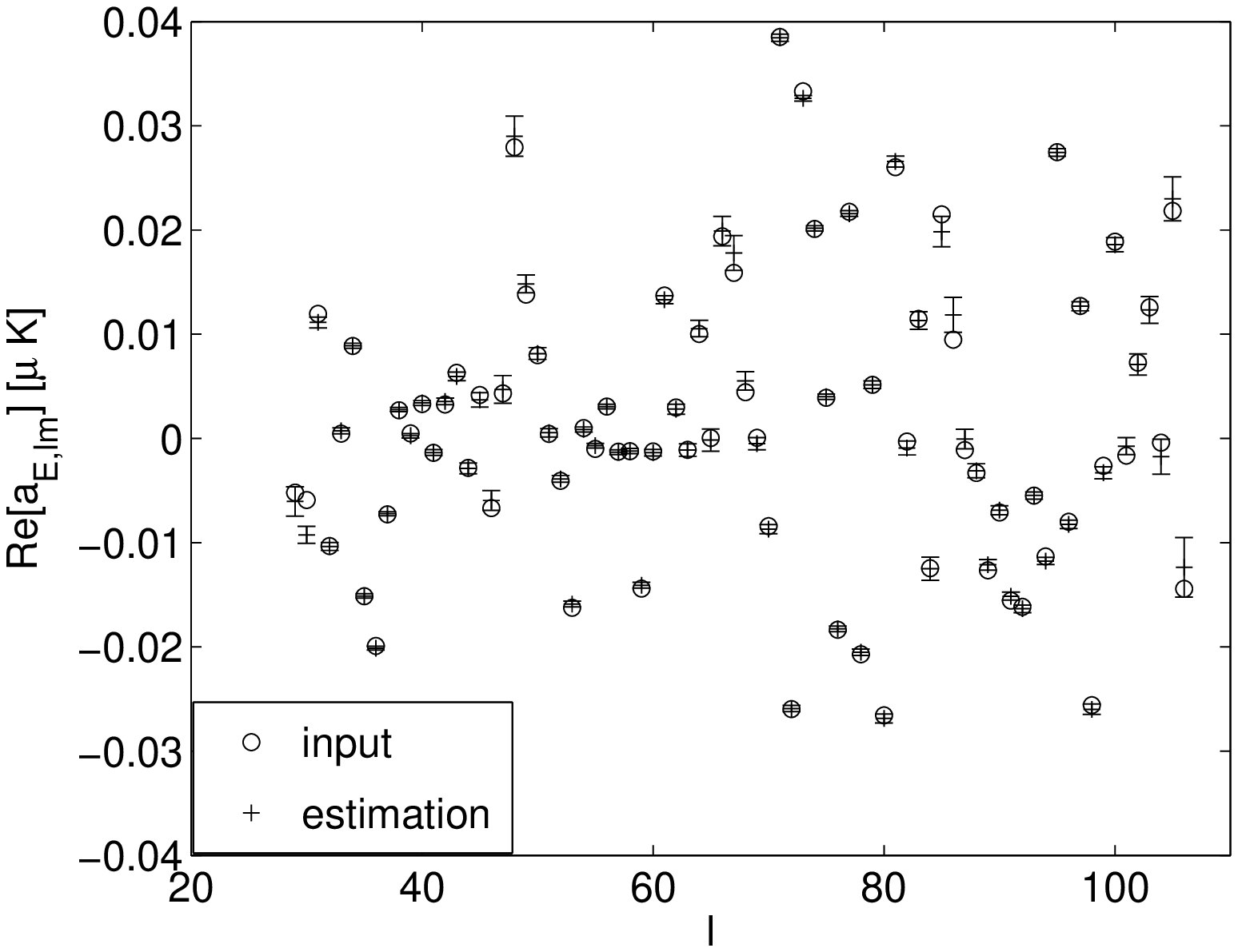}
\caption{estimated $\mathrm {Re}[a_{E,lm}]$ (m=29)}
\label{E_r-noise}
\end{figure}

\begin{figure}
\centering
\centering\includegraphics[scale=.5]{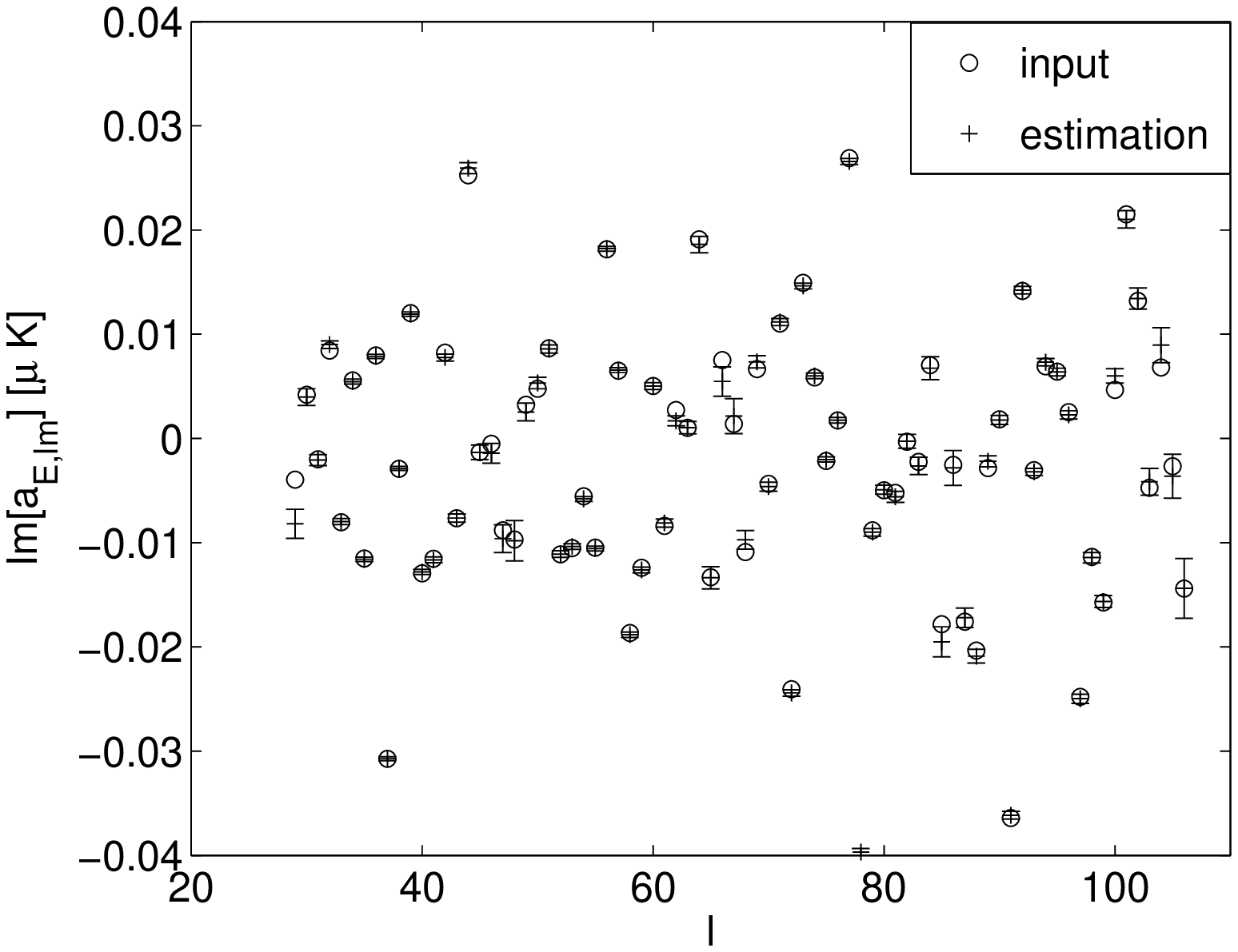}
\caption{estimated $\mathrm {Im}[a_{E,lm}]$ (m=29)}
\label{E_i-noise}
\end{figure}

\begin{figure}
\centering
\centering\includegraphics[scale=.5]{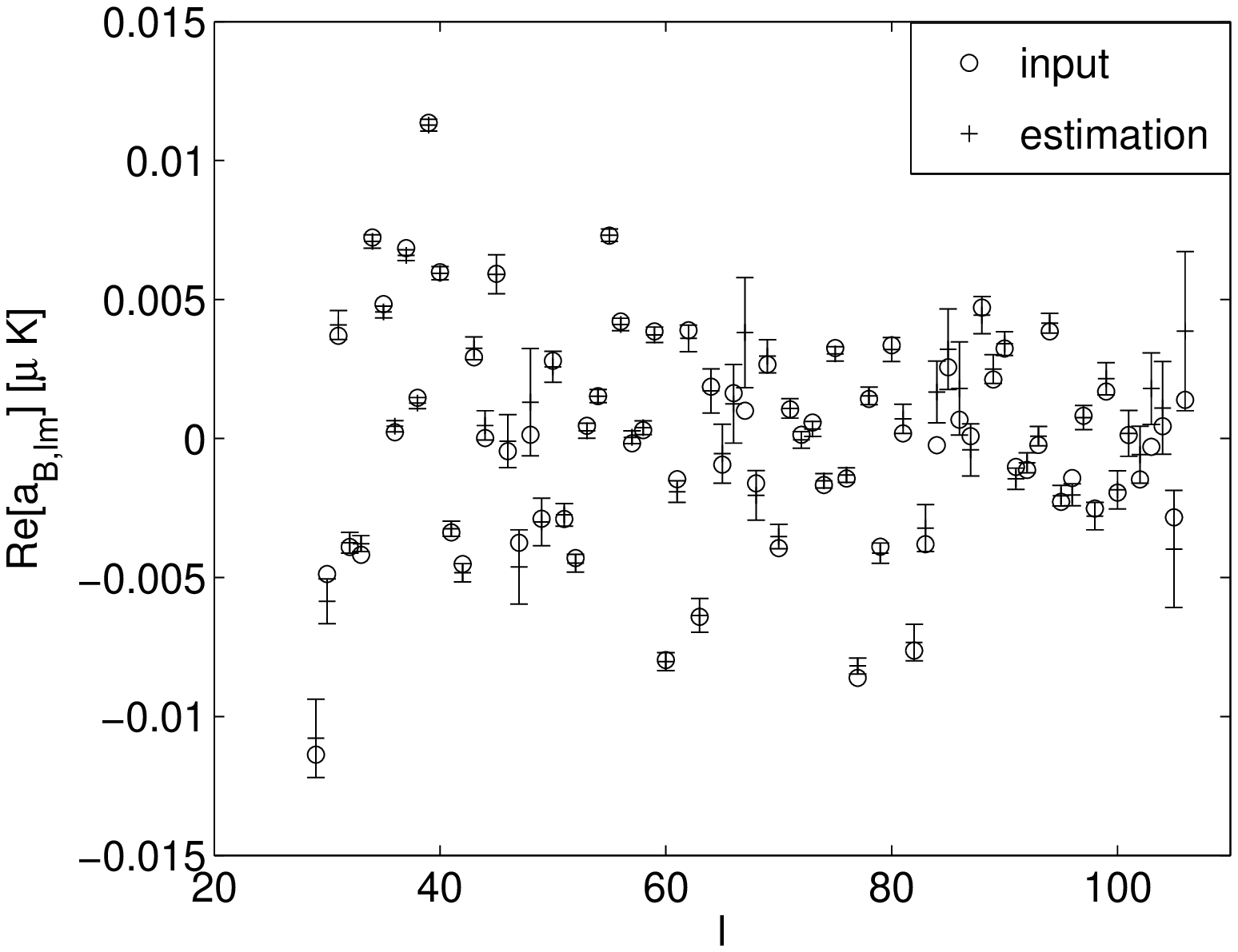}
\caption{estimated $\mathrm {Re}[a_{B,lm}]$ (m=29)}
\label{B_r-noise}
\end{figure}

\begin{figure}
\centering
\centering\includegraphics[scale=.5]{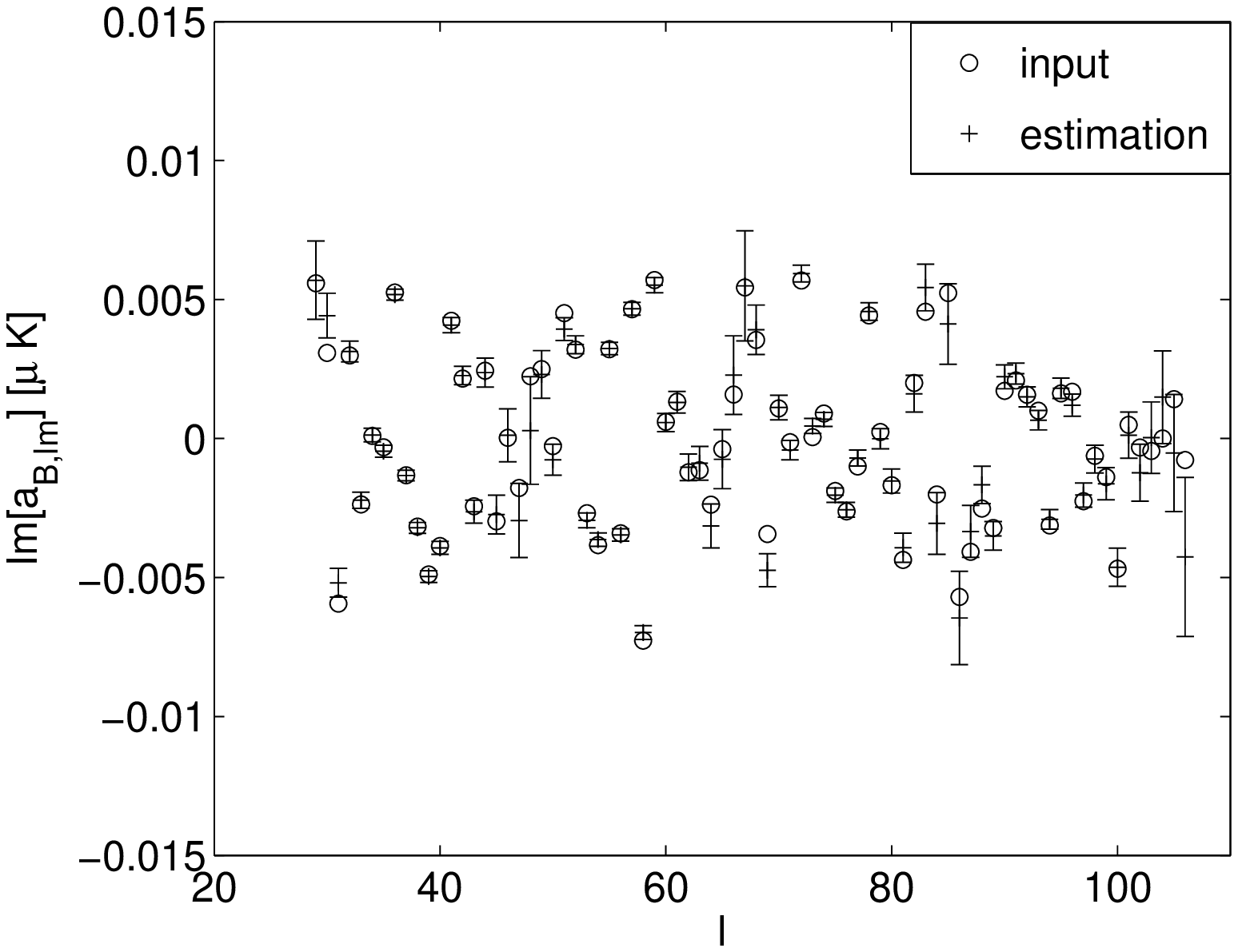}
\caption{estimated $\mathrm {Im}[a_{B,lm}]$ (m=29)}
\label{B_i-noise}
\end{figure}

With Eq. \ref{maximum_likelihood}, we have estimated spherical harmonic coefficients $a_{2,lm}$ $(29 \le l \le 48)$ from $B_1$ visibilities, $(48 \le l \le 67)$ from $B_2$, $(67 \le l \le 86)$ from $B_3$, and $(86 \le l \le 106)$ from $B_4$. 
From the estimated $a_{2,lm}$, we have obtained $a_{E,lm}$ and $a_{B,lm}$ via Eq. \ref{E_decomposition} and \ref{B_decomposition}.
Estimated $a_{E,lm}$ and $a_{B,lm}$ $(\;m=29)$ are shown together with the input $a_{E,lm}$ and $a_{B,lm}$ value in Fig. 
\ref{E_r-noise}, \ref{E_i-noise}, \ref{B_r-noise}, and \ref{B_i-noise}.
These are representative of other $m$ modes.  
The $1-\sigma$ errors via Eq. \ref{error_covariance} are indicated by vertical error bars in Fig. \ref{E_r-noise}, \ref{E_i-noise}, \ref{B_r-noise}, and \ref{B_i-noise}.
As shown in Fig. \ref{window_function}, the window function of the simulated observation
have several troughs, where $1-\sigma$ errors are expected to be large. It is seen that $1-\sigma$ errors are significant on the multipoles corresponding to the troughs of the window function.
By averaging the magnitude of 1-$\sigma$ error on $a_{E,lm}$ and $a_{B,lm}$,
we found that they are in same magnitude within 1\%.
It is not suprising, considering the expression for the variance of
$\Delta a_{E,lm}$ and $\Delta a_{B,lm}$, which are shown in Eq. \ref{E_variance} and \ref{B_variance}.

\section{DISCUSSION}
Visibilities associated with the CMB polarization can be expressed as a linear sum of  spherical harmonic coefficients $a_{\pm 2,lm}$ of the CMB polarization. The linear weight for $a_{\pm 2,lm}$ depends on the observational configuration, and spherical harmonic number $l,m$. Since an interferometer is sensitive over a finite range of multipoles.
The spherical harmonic coefficients ($l_0\le l\le l_1$) can be determined by fitting $a_{\pm 2,lm}$ for visibilities of various observational configuration.
Once $a_{\pm2,lm}$ are determined, $a_{E,lm}$ and $a_{B,lm}$ are easily obtained via Eq. \ref{E_decomposition} and \ref{B_decomposition}.
The linear weights $b_{\pm 2,lm}$, which map visibilities to  spherical harmonic space, can be computed fast with the aid of the methods discussed in Appendix \ref{computing_b2lm}. 
The best-fit value of $a_{2,lm}$ for given visibilities can be found via Eq. \ref{maximum_likelihood}. It is $\mathcal O(\mathsf m^3)$ process, where $\mathsf m$ is the number of $a_{2,lm}$ to be determined.
Scaling the time taken for the simulated observation says this formalism is computationally feasible for interferometric observation up to multipoles as high as $l\sim 1500$.

Since the formalism introduced in this paper is developed for a satelite-based interferometric observations of all-sky polarization such as the EPIC \citep{Timbie:Irvine}, the antenna pointings in the simulated observation are made over full-sky. Even when antenna pointings are made within a fraction of sky, the matrix  $\mathbf b$ in Eq. \ref{Vba} does not become singular, as far as the fraction of sky is big enough in relative to the angular scales the interferometer is sensitive to.
 
$a_{\pm2,lm}$ determined by Eq. \ref{maximum_likelihood} contain foreground contamination
like pixellized maps.
With different frequency spectral behavior of foregrounds from the CMB, the foreground contribution can be separated with the multi-frequency data from the CMB down to residual level, which is limited by frequency spectral incoherence and 
knowledge of polarized foregrounds \citep{Tegmark:Foreground,Tegmark:Forecasts}.

There are several sources for E/B mode mixing.
Aliasing due to finite pixel size leads to E/B mode mixing
at high multipoles and limited sky coverage does at low multipoles.
Interferometer observations enable targeting a specific range of multipoles. E/B mode mixing at low multipoles due to limited sky coverage can be made insignificant by designing interferometers to be insensitive to anisotropy at low multipoles. 
Since the formalism reconstructs $a_{E,lm}$ and $a_{B,lm}$ directly from spherical harmonic space, E/B mixing due to pixellization and ambiguity of E/B mode over the mosaic 
are insignificant.

We choose the multipole range $l_0$ and  $l_1$ to be the first multipoles, where the window function drops below 1\% of its peak value. There are residual contribution from
$a_{\pm2,lm}$ ($l< l_0$, $l> l_1$).
These residuals are another source of estimation error in addition to instrument noise.
We can modify noise covariance matrix of Eq. \ref{maximum_likelihood} to include $\sum_l C_l \mathbf W_{l,ij}$, where $C_l$ and $\mathbf W_{l,ij}$ is power spectra and window functions at out-of-bound multipoles.
It improves the estimation error due to residuals by giving more weights to the visibilities of less contribution from residuals. But it increases computational load, by making a total noise covariance matrix
non-diagonal, while an instrument noise covariance matrix is diagonal to a good approximation \citep{Hobson:maximum_likelihood}.

We have presented a formalism to reconstruct spherical harmonics of the CMB polarization directly from interferometer observations. The formalism takes advantage of the fact that an interferometer directly probes the Fourier components of sky pattern, and the relation between a Fourier component and spin $\pm2$ spherical harmonics.
\section{ACKNOWLEDGMENTS}
The author thanks Gregory Tucker, Peter Timbie, Emory Bunn, Andrei Korotkov and Carolina Calderon for useful discussions.
He thanks Douglas Scott for the hospitality during the visit to UBC.
He thanks anonymous referees for thorough reading and helpful comments, which led to significant improvements in the paper.
\bibliographystyle{plainnat}
\bibliography{/home/tac/jkim/Documents/bibliography}

\begin{thebibliography}{33}
\expandafter\ifx\csname natexlab\endcsname\relax\def\natexlab#1{#1}\fi
\expandafter\ifx\csname url\endcsname\relax
  \def\url#1{{\tt #1}}\fi

\bibitem[Arfken and Weber(2000)]{Arfken}
George~B. Arfken and Hans~J. Weber.
\newblock {\em Mathematical Methods for Physicists}.
\newblock Academic Press, San Diego, CA USA, 5th edition, 2000.

\bibitem[Barkats et~al.(2005)Barkats, Bischoff, Farese, Fitzpatrick, Gaier,
  Gundersen, Hedman, Hyatt, McMahon, Samtleben, Staggs, Vanderlinde, and
  Winstein]{CAPMAP:polarization}
D.~Barkats, C.~Bischoff, P.~Farese, L.~Fitzpatrick, T.~Gaier, J.~O. Gundersen,
  M.~M. Hedman, L.~Hyatt, J.~J. McMahon, D.~Samtleben, S.~T. Staggs,
  K.~Vanderlinde, and B.~Winstein.
\newblock First measurements of the polarization of the cosmic microwave
  background radiation at small angular scales from capmap.
\newblock {\em Astrophys. J. Lett.}, 619:\penalty0 127, 2005.

\bibitem[Challinor and Lewis(2005)]{CMB_Lensing:correlation}
Anthony Challinor and Antony Lewis.
\newblock Lensed {CMB} power spectra from all-sky correlation functions.
\newblock {\em Phys. Rev. D}, 71, 2005.

\bibitem[Dodelson(2003)]{Modern_Cosmology}
Scott Dodelson.
\newblock {\em Modern Cosmology}.
\newblock Academic Press, 2nd edition, 2003.

\bibitem[Efstathiou(2006)]{hybrid_estimation}
G.~Efstathiou.
\newblock Hybrid estimation of {CMB} polarization power spectra.
\newblock {\em Mon. Not. R. Astron. Soc.}, 370:\penalty0 343, 2006.

\bibitem[Gorski et~al.(2005)Gorski, Hivon, Banday, Wandelt, Hansen, Reinecke,
  and Bartelman]{HEALPix:framework}
K.~M. Gorski, E.~Hivon, A.~J. Banday, B.~D. Wandelt, F.~K. Hansen, M.~Reinecke,
  and M.~Bartelman.
\newblock {HEALPix} -- a framework for high resolution discretization, and fast
  analysis of data distributed on the sphere.
\newblock {\em Astrophys. J.}, 622:\penalty0 759, 2005.

\bibitem[Hinshaw and et~al.(2007)]{WMAP:3yr_TT}
G.~Hinshaw and et~al.
\newblock Three-year {Wilkinson Microwave Anisotropy Probe} ({WMAP})
  observations: Temperature analysis.
\newblock {\em Astrophys.J.Suppl.}, 170:\penalty0 288, 2007.
\newblock http://lambda.gsfc.nasa.gov.

\bibitem[Kamionkowski et~al.(1997)Kamionkowski, Kosowsky, and
  Stebbins]{Kamionkowski:Flm}
Marc Kamionkowski, Arthur Kosowsky, and Albert Stebbins.
\newblock Statistics of cosmic microwave background polarization.
\newblock {\em Phys. Rev. D}, 55:\penalty0 7368, 1997.

\bibitem[Korotkov et~al.(2006)Korotkov, Kim, Tucker, Gault, Hyland, Malu,
  Timbie, Bunn, Bierman, Keating, Murphy, O'Sullivan, Ade, Calderon, and
  Piccirillo]{SPIE:MBI}
Andrei~L. Korotkov, Jaiseung Kim, Gregory~S. Tucker, Amanda Gault, Peter
  Hyland, Siddharth Malu, Peter~T. Timbie, Emory~F. Bunn, Evan Bierman, Brian
  Keating, Anthony Murphy, Créidhe O'Sullivan, Peter~A. Ade, Carolina
  Calderon, and Lucio Piccirillo.
\newblock The millimeter-wave bolometric interferometer.
\newblock In Jonas Zmuidzinas, Wayne~S. Holland, Stafford Withington, and
  William~D. Duncan, editors, {\em Millimeter and Submillimeter Detectors and
  Instrumentation for Astronomy III: Proceedings of the SPIE}, volume 6275,
  Bellingham, WA USA, Jun 2006. SPIE, The International Society for Optical
  Engineering.

\bibitem[Kovac et~al.(2002)Kovac, Leitch, Pryke, Carlstrom, Halverson, and
  Holzapfel]{DASI:data}
J.~Kovac, E.~M. Leitch, C.~Pryke, J.~E. Carlstrom, N.~W. Halverson, and W.~L.
  Holzapfel.
\newblock Detection of polarization in the cosmic microwave background using
  {DASI}.
\newblock {\em Nature}, 420:\penalty0 772, 2002.

\bibitem[Kraus(1986)]{Kraus:Radio_Astronomy}
J.~Kraus.
\newblock {\em Radio Astronomy}.
\newblock Cygnus-Quasar Books, Powell, Ohio USA, 2nd edition, 1986.

\bibitem[Lawson(2006)]{Lawson:Interferometry}
P.~Lawson, editor.
\newblock {\em Principles of Long Baseline Stellar Interferometry}.
\newblock Wiley-Interscience, Mississauga, Ontario Canada, 2006.

\bibitem[Leitch et~al.(2002)Leitch, Kovac, Pryke, Reddall, Sandberg, Dragovan,
  Carlstrom, Halverson, and Holzapfel]{DASI:instrument}
E.~M. Leitch, J.~M. Kovac, C.~Pryke, B.~Reddall, E.~S. Sandberg, M.~Dragovan,
  J.~E. Carlstrom, N.~W. Halverson, and W.~L. Holzapfel.
\newblock Measurement of polarization with the degree angular scale
  interferometer.
\newblock {\em Nature}, 420:\penalty0 763, 2002.

\bibitem[Leitch et~al.(2005)Leitch, Kovac, Halverson, Carlstrom, Pryke, and
  Smith]{DASI:3yr}
Erik~M. Leitch, J.~M. Kovac, N.~W. Halverson, J.~E. Carlstrom, C.~Pryke, and
  M.~W.~E. Smith.
\newblock {DASI} three-year cosmic microwave background polarization results.
\newblock {\em Astrophys. J.}, 624:\penalty0 10--20, 2005.

\bibitem[Lewis et~al.(2000)Lewis, Challinor, and Lasenby]{CAMB}
Antony Lewis, Anthony Challinor, and Anthony Lasenby.
\newblock Efficient computation of {CMB} anisotropies in closed {FRW} models.
\newblock {\em Astrophys. J.}, 538:\penalty0 473, 2000.
\newblock http://camb.info/.

\bibitem[Montroy et~al.(2006)Montroy, Ade, Bock, Bond, Borrill, Boscaleri,
  Cabella, Contaldi, Crill, de~Bernardis, Gasperis, de~Oliveira-Costa, Troia,
  di~Stefano, Hivon, Jaffe, Kisner, Jones, Lange, Masi, Mauskopf, MacTavish,
  Melchiorri, Natoli, Netterfield, Pascale, Piacentini, Pogosyan, Polenta,
  Prunet, Ricciardi, Romeo, Ruhl, Santini, Tegmark, Veneziani, and
  Vittorio]{BOOMERANG:Polarization}
T.~E. Montroy, P.~A.~R. Ade, J.~J. Bock, J.~R. Bond, J.~Borrill, A.~Boscaleri,
  P.~Cabella, C.~R. Contaldi, B.~P. Crill, P.~de~Bernardis, G.~De Gasperis,
  A.~de~Oliveira-Costa, G.~De Troia, G.~di~Stefano, E.~Hivon, A.~H. Jaffe,
  T.~S. Kisner, W.~C. Jones, A.~E. Lange, S.~Masi, P.~D. Mauskopf, C.~J.
  MacTavish, A.~Melchiorri, P.~Natoli, C.~B. Netterfield, E.~Pascale,
  F.~Piacentini, D.~Pogosyan, G.~Polenta, S.~Prunet, S.~Ricciardi, G.~Romeo,
  J.~E. Ruhl, P.~Santini, M.~Tegmark, M.~Veneziani, and N.~Vittorio.
\newblock A measurement of the {CMB} <ee〉 spectrum from the 2003 flight of
  {BOOMERANG}.
\newblock {\em Astrophys. J.}, 647:\penalty0 813, 2006.

\bibitem[M.P.Hobson and Maisinger(2002)]{Hobson:maximum_likelihood}
M.P.Hobson and Klaus Maisinger.
\newblock Maximum-likelihood estimation of the {CMB} power spectrum from
  interferometer observations.
\newblock {\em Mon. Not. R. Astron. Soc.}, 334:\penalty0 569, 2002.

\bibitem[Muciaccia et~al.(1997)Muciaccia, Natoli, and
  Vittorio]{FastSphericalHarmonics}
P.~F. Muciaccia, P.~Natoli, and N.~Vittorio.
\newblock Fast spherical harmonic analysis: A quick algorithm for generating
  and/or inverting full-sky, high-resolution cosmic microwave background
  anisotropy maps.
\newblock {\em Astrophys. J. Lett.}, 488:\penalty0 123523, 1997.

\bibitem[Page et~al.(2006)Page, Hinshaw, Komatsu, Nolta, Spergel, Bennett,
  Barnes, Bean, Dore', Halpern, Hill, Jarosik, Kogut, Limon, Meyer, Odegard,
  Peiris, Tucker, Verde, Weiland, Wollack, and Wright]{WMAP:polarization}
L.~Page, G.~Hinshaw, E.~Komatsu, M.~R. Nolta, D.~N. Spergel, C.~L. Bennett,
  C.~Barnes, R.~Bean, O.~Dore', M.~Halpern, R.~S. Hill, N.~Jarosik, A.~Kogut,
  M.~Limon, S.~S. Meyer, N.~Odegard, H.~V. Peiris, G.~S. Tucker, L.~Verde,
  J.~L. Weiland, E.~Wollack, and E.~L. Wright.
\newblock Three year wilkinson microwave anisotropy probe ({WMAP})
  observations: Polarization analysis.
\newblock {\em Accepted by ApJ}, 2006.

\bibitem[Park and Ng(2004)]{Park:EB_separation}
Chan-Gyung Park and Kin-Wang Ng.
\newblock {E/B} separation in {CMB} interferometry.
\newblock {\em Astrophys. J.}, 609:\penalty0 15, 2004.

\bibitem[Park et~al.(2003)Park, Ng, Park, Liu, and Umetsu]{CMB:strategy}
Chan-Gyung Park, Kin-Wang Ng, Changbom Park, Guo-Chin Liu, and Keiichi Umetsu.
\newblock Observational strategies of {CMB} temperature and polarization
  interferometry experiments.
\newblock {\em Astrophys. J.}, 589:\penalty0 67, 2003.

\bibitem[Press et~al.(1992)Press, Flannery, Teukolsky, and
  Vetterling]{Numerical_Recipe_C}
William~H. Press, Brian~P. Flannery, Saul~A. Teukolsky, and William~T.
  Vetterling.
\newblock {\em Numerical Recipes in C : The Art of Scientific Computing}.
\newblock Cambridge University Press, 2nd edition, 1992.

\bibitem[Readhead et~al.(2004)Readhead, Myers, Pearson, Sievers, Mason,
  Contaldi, Bond, Bustos, Altamirano, Achermann, Bronfman, Carlstrom,
  Cartwright, Casassus, Dickinson, Holzapfel, Kovac, Leitch, May, Padin,
  Pogosyan, Pospieszalski, Pryke, Reeves, Shepherd, and
  Torres]{CBI:polarization}
A.~C.~S. Readhead, S.~T. Myers, T.~J. Pearson, J.~L. Sievers, B.~S. Mason,
  C.~R. Contaldi, J.~R. Bond, R.~Bustos, P.~Altamirano, C.~Achermann,
  L.~Bronfman, J.~E. Carlstrom, J.~K. Cartwright, S.~Casassus, C.~Dickinson,
  W.~L. Holzapfel, J.~M. Kovac, E.~M. Leitch, J.~May, S.~Padin, D.~Pogosyan,
  M.~Pospieszalski, C.~Pryke, R.~Reeves, M.~C. Shepherd, and S.~Torres.
\newblock Polarization observations with the cosmic background imager.
\newblock {\em Science}, 306:\penalty0 836, 2004.
\newblock http://www.astro.caltech.edu/~tjp/CBI/press2/index.html.

\bibitem[Rohlfs and Wilson(2003)]{Tools_Radio_Astronomy}
K.~Rohlfs and T.~L. Wilson.
\newblock {\em Tools of Radio Astronomy}.
\newblock Springer-Verlag, New York, NY USA, 4th edition, 2003.

\bibitem[Tauber(2000)]{Planck:sensitivity}
J.~A. Tauber.
\newblock The {Planck} mission: Overview and current status.
\newblock {\em Astrophysical Letters and Communications}, 37:\penalty0 145,
  2000.
\newblock http://planck.esa.int.

\bibitem[Tegmark and Efstathiou(1996)]{Tegmark:Foreground}
M.~Tegmark and G.~Efstathiou.
\newblock A method for subtracting foregrounds from multi-frequency {CMB} sky
  maps.
\newblock {\em Mon. Not. R. Astron. Soc.}, 281:\penalty0 1297, 1996.

\bibitem[Tegmark et~al.(2000)Tegmark, Eisenstein, Hu, and
  de~Oliveira-Costa]{Tegmark:Forecasts}
Max Tegmark, Daniel~J. Eisenstein, Wayne Hu, and Angelica de~Oliveira-Costa.
\newblock Foregrounds and forecasts for the cosmic microwave background.
\newblock {\em Astrophys. J.}, 530:\penalty0 133, 2000.

\bibitem[Thompson et~al.(2001)Thompson, Moran, and
  Swenson]{Thompson:interferometer}
A.~Richard Thompson, James~M. Moran, and George~W. Swenson, Jr.
\newblock {\em Interferometry and synthesis in radio astronomy}.
\newblock Wiley-Intescience Publication, Mississauga, Ontario Canada, 2nd
  edition, 2001.

\bibitem[Timbie et~al.(2006)Timbie, Tucker, Ade, Ali, Bierman, Bunn, Calderon,
  Gault, Hyland, Keating, Kim, Korotkov, Malu, Mauskopf, Murphy, O'Sullivan,
  Piccirillo, and Wandelt]{Timbie:Irvine}
P.~T. Timbie, G.~S. Tucker, P.~A.~R. Ade, S.~Ali, E.~Bierman, E.~F. Bunn,
  C.~Calderon, A.~C. Gault, P.~O. Hyland, B.~G. Keating, J.~Kim, A.~Korotkov,
  S.~Malu, P.~Mauskopf, J.~A. Murphy, C.~O'Sullivan, L.~Piccirillo, and B.~D.
  Wandelt.
\newblock The einstein polarization interferometer for cosmology ({EPIC}) and
  the millimeterwave bolometric interferometer ({MBI}).
\newblock {\em New Astronomy Reviews}, 50:\penalty0 999--1008, 2006.

\bibitem[Tucker et~al.(2003)Tucker, Kim, Timbie, Alib, Piccirilloc, and
  Calderon]{Tucker:Bolometric_interferometry}
G.~S. Tucker, J.~Kim, P.~Timbie, S.~Alib, L.~Piccirilloc, and C.~Calderon.
\newblock Bolometric interferometry: the millimeter-wave bolometric
  interferometer.
\newblock {\em New Astronomy Reviews}, 47:\penalty0 1173--1176, 2003.

\bibitem[Zaldarriaga(1998{\natexlab{a}})]{Zaldarriaga:Polarization_Exp}
M.~Zaldarriaga.
\newblock {CMB} polarization experiments.
\newblock {\em Astrophys. J.}, 503:\penalty0 1, 1998{\natexlab{a}}.

\bibitem[Zaldarriaga and Seljak(1997)]{Seljak-Zaldarriaga:Polarization}
M.~Zaldarriaga and U.~Seljak.
\newblock An all-sky analysis of polarization in the microwave background.
\newblock {\em Phys. Rev. D}, 55:\penalty0 1830, 1997.

\bibitem[Zaldarriaga(1998{\natexlab{b}})]{Zaldarriaga:thesis}
Matias Zaldarriaga.
\newblock {\em Fluctuations in the Cosmic Microwave Background}.
\newblock PhD thesis, MIT, 1998{\natexlab{b}}.

\end{thebibliography}
\appendix

\section{CMB polarization basis vectors and antenna coordinate}
\label{relation}
In all-sky analysis, the CMB polarization at the angular coordinate ($\theta,\phi$) 
are measured in the local reference frame whose axises are 
($\mathbf {\hat e}_{\theta}$, $\mathbf {\hat e}_{\phi}$, $\mathbf {\hat e}_{r}$).
Let's call this coordinate frame `the local CMBP frame' from now on.
Consider the polarization observation of antenna pointing ($\theta_A$,$\phi_A$). 
A global coordinate frame coincides with the antenna coordinate by Euler rotations $\textbf{R}_y(\theta_A)\,\textbf{R}_z(\phi_A)$.
Since a global coordinate frame coincides with the local CMBP frame by Euler rotations 
$\textbf{R}_y(\theta)\,\textbf{R}_z(\phi)$, the Euler Rotations $\textbf{R}_z(\gamma)\,\textbf{R}_y(\beta)\,\textbf{R}_z(\alpha)$
coincides the antenna coordinate frame with the local CMBP frame,
where $\textbf{R}_z(\gamma)\,\textbf{R}_y(\beta)\,\textbf{R}_z(\alpha)
\,\textbf{R}_y(\theta_A)\,\textbf{R}_z(\phi_A)=\textbf{R}_y(\theta)\,\textbf{R}_z(\phi)$.
Therefore, the local CMBP frame is in rotation from the antenna coordinate by the Euler angles $(\alpha,\beta,\gamma)$ as follows:   
\begin{eqnarray*}
\alpha&=&\tan^{-1}\left[\frac{\sin\theta\sin(\phi-\phi_A)}{\sin\theta\cos\theta_A\cos(\phi-\phi_A)-\cos\theta\sin\theta_A}\right],\\
\beta&=&\cos^{-1}\left[\cos\theta\cos\theta_A+\sin\theta\sin\theta_A\cos(\phi-\phi_A)\right],\\
\gamma&=&\tan^{-1}\left[\frac{\sin\theta_A\sin(\phi-\phi_A)}{-\sin\theta\cos\theta_A+\cos\theta\sin\theta_A\cos(\phi-\phi_A)} \right],
\end{eqnarray*}
where the Euler angles $(\alpha,\beta,\gamma)$ can be obtained from
$\textbf{R}_z(\gamma)\,\textbf{R}_y(\beta)\,\textbf{R}_z(\alpha)
=\textbf{R}_y(\theta)\,\textbf{R}_z(\phi)\,\textbf{R}^{-1}_z(\phi_A)\,\textbf{R}^{-1}_y(\theta_A)$.
In most CMB polarization experiments, where polarizers are attached to the other side of feedhorns, incoming rays go through polarizers after feedhorns. After passing through a feedhorn, an incoming off-axis ray becomes an on-axis ray.
Then the local CMBP frame of the ray after the feedhorn system is simply in azimuthal rotation 
$\alpha+\gamma$ from the antenna coordinate.
Therefore, $\Phi$ in Eq. \ref{Phi} is
\begin{eqnarray*}
\Phi&=&\tan^{-1}\left[\frac{\sin\theta\sin(\phi-\phi_A)}{\sin\theta\cos\theta_A\cos(\phi-\phi_A)-\cos\theta\sin\theta_A}\right]\\
&&+\tan^{-1}\left[\frac{\sin\theta_A\sin(\phi-\phi_A)}{-\sin\theta\cos\theta_A+\cos\theta\sin\theta_A\cos(\phi-\phi_A)} \right].
\end{eqnarray*}

\section{Computing linear weights}
\label{computing_b2lm}
${b_{\pm2,lm}}$ needs be computed to determine $a_{2,lm}$ ($l_0\le l\le l_1$)
via Eq. \ref{maximum_likelihood}. 
It can be computed in the baseline coordinate  
where a baseline coincides with the $x$ axis of the coordinate.
In computing ${b_{\pm2,lm}}$ in the baseline coordinate, spin $\pm2$ spherical harmonics, which are defined in the global coordinate, are related to spin $\pm2$ spherical harmonics in the baseline coordinate with a rotation matrix \citep{CMB_Lensing:correlation}. 
Let's choose Galactic coordinate as the global reference coordinate for the CMB.
Consider the polarization observation, whose antenna pointing is in the direction of Galactic coordinate ($\ell$,$b$). The baseline is assumed to be on the aperture plane.
Then, the baseline coordinate is the coordinate system rotated from Galactic coordinate by Euler rotation $\mathcal R(\ell,\frac{\pi}{2}-b,\phi_\mathbf u)$, where $\phi_\mathbf u$ is the rotation around the axis of antenna pointing.  
Computed in the baseline coordinate, $b_{\pm 2,lm}$ are
\begin{eqnarray}
b_{2,lm}&=&\int d\nu\,f(\nu)\int^\pi_{0} \mathrm d(\theta^\prime)\sin\theta^\prime A(\theta')\nonumber\\
&&\times\int^{2\pi}_0 \mathrm d\phi'\:\sum_{m'}\mathsf D^l_{m'm}(\ell,\frac{\pi}{2}-b,\phi_{\mathbf u})\;{}_{2}Y_{lm'}(\theta',\phi')\nonumber\\
&&\times \mathrm e^{\mathrm i (2\pi u \sin\theta^\prime\cos\phi'-2(\psi-\phi_\mathbf u)+2\phi')}\nonumber\\
&=&-\mathrm e^{-\mathrm i 2\psi}\sqrt{\pi(2l+1)}\nonumber\\
&&\times\sum_{m'}
\mathrm e^{\mathrm i(m^\prime\frac{\pi}{2}+2\phi_{\mathbf u})}\;\mathsf D^l_{m'm}(\ell,\frac{\pi}{2}-b,\phi_{\mathbf u})\nonumber\\
&&\times \int^1_{-1} \mathrm d(\cos\theta')A(\theta')[F_{1,lm'}(\theta')+F_{2,lm'}(\theta')]\nonumber\\
&&\times \int d\nu\,f(\nu)J_{m'+2}(2\pi \sin\theta^\prime u),\label{b_2lm_baseline}
\end{eqnarray}
\begin{eqnarray}
b_{-2,lm}&=&\int d\nu\,f(\nu)\int^1_{-1} \mathrm d(\cos\theta^\prime)A(\theta')\nonumber\\
&&\times \int^{2\pi}_0 \mathrm d\phi'\: \sum_{m'}\mathsf D^l_{m'm}(\ell,\frac{\pi}{2}-b,\phi_{\mathbf u})\;{}_{-2}Y_{lm'}(\theta',\phi')\nonumber\\
&&\times \mathrm e^{\mathrm i (2\pi u \sin\theta^\prime \cos\phi^\prime+2(\psi-\phi_{\mathbf u})-2\phi')}\nonumber\\
&=&-\mathrm e^{\mathrm i 2\psi}\sqrt{\pi(2l+1)}\nonumber\\
&&\times\sum_{m'}\mathrm e^{\mathrm i(m^\prime\frac{\pi}{2}-2\phi_{\mathbf u})}\;\mathsf D^l_{m'm}(\ell,\frac{\pi}{2}-b,\phi_{\mathbf u})\nonumber\\
&&\times \int^1_{-1} \mathrm d(\cos\theta')A(\theta')[F_{1,lm'}(\theta')-F_{2,lm'}(\theta')]\nonumber\\
&&\times \int d\nu\,f(\nu)J_{m'-2}(2\pi \sin\theta^\prime u),\label{b_-2lm_baseline}
\end{eqnarray}
where $J_{n}$ is the $n$th order ordinary Bessel function.
It turns out that computing $b_{\pm2,lm}$ via Eq. \ref{b_2lm_baseline} and \ref{b_-2lm_baseline} suffers from serious numerical precision problem especially for high multipole ($l>60$), which 
are due to machine floating-point rounding error occurring in the multiplication with the rotation matrix $\mathsf D^l_{m'm}(\mathcal R)$.
Besides the numerical precision problem, the huge time required for computing the rotation matrix $\mathsf D^l_{m'm}(\mathcal R)$ makes 
Eq. \ref{b_2lm_baseline} and Eq. \ref{b_-2lm_baseline} lose most of merits 
gained by the availability of analytic integration over azimuthal angle.
For these reasons,
in the simulated observation of \S \ref{simulation} we computed $b_{\pm2,lm}$ in a fixed CMB frame with Eq. \ref{b_2lm} and \ref{b_-2lm}.
In Eq. \ref{b_2lm} and \ref{b_-2lm}, we have rearranged the order of integration and replaced the integration over continuum with sum over finite elements, which are as follows:
\begin{eqnarray}
b_{2,lm}&=&\sqrt{\frac{2l+1}{4\pi}}\mathrm e^{-\mathrm i 2\psi}\Delta \theta \Delta \phi \Delta \nu\label{b_2lm_fast}\\
&&\times\sum_i \sin\theta_i
(F_{1,lm}(\theta_i)+F_{2,lm}(\theta_i))\nonumber\\
&&\times \sum_j 
\mathrm e^{\mathrm i m\phi_j}\mathrm e^{\mathrm i 2\Phi(\hat {\mathbf n})}\sum_k f(\nu_k)A(\hat {\mathbf n},\hat {\mathbf n}_A)
\mathrm e^{\mathrm i(2\pi\mathbf u_k\cdot \hat {\mathbf n})},\nonumber\\
b_{-2,lm}&=&\sqrt{\frac{2l+1}{4\pi}}\mathrm e^{\mathrm i 2\psi}\Delta \theta \Delta \phi \Delta \nu\label{b_-2lm_fast}\\
&&\times \sum_i \sin\theta_i
(F_{1,lm}(\theta_i)-F_{2,lm}(\theta_i))\nonumber\\
&&\times \sum_j 
\mathrm e^{\mathrm i m\phi_j}\mathrm e^{-\mathrm i 2\Phi(\hat {\mathbf n})}\sum_k f(\nu_k)A(\hat {\mathbf n},\hat {\mathbf n}_A)
\mathrm e^{\mathrm i(2\pi\mathbf u_k\cdot \hat {\mathbf n})}.\nonumber
\end{eqnarray} 
Legendre functions in $F_{1,lm}(\theta)$ and $F_{2,lm}(\theta)$ are computed with the following recurrence relation
for $x=\cos\theta$ \citep{Numerical_Recipe_C}:
\begin{eqnarray*}
(l-m)P^m_l&=&x(2l-1)P^m_{l-1}-(l+m-1)P^m_{l-2},\\
P^m_m&=&(-1)^m(2m-1)!!(1-x^2)^{m.2},\\
P^m_{m+1}&=&x(2m+1)P^m_m.
\end{eqnarray*}
When an interferometer is sensitive to multipole range $l_{0}\le l\le l_{1}$, 
$b_{\pm2,lm}$ of $l$ up to $l_{1}$ should be computed. 
As shown in Eq. \ref{2Ylm}, Eq. \ref{-2Ylm}, Eq. \ref{F1lm} and Eq. \ref{F2lm}, 
${}_{\pm2}Y_{lm}$ is the product of Legendre functions and $e^{i m\phi}$.
Legendre functions of multipole $l$ varies on angular scale down to  $\theta\approx180^\circ/l$.
With Nyquist-Shannon Sampling theorem \citep{Numerical_Recipe_C}, the integration over $\theta$ should be done with $\Delta \theta<\frac{180^{\circ}}{2l_{1}}$. 
Since `$e^{i m\phi}$' consisting of ${}_{\pm2}Y_{lm}$ is periodic over $\phi=360^\circ/m$,
the integration over $\phi$ should be also done with $\Delta \phi<\frac{360^{\circ}}{2l_{1}}$.

The summation of index $j$ below, which is part of Eq. \ref{b_2lm_fast} and \ref{b_-2lm_fast}, is equivalent to discrete Fourier Transform:
\[\sum_j \mathrm e^{\mathrm i m\phi_j}\left[\mathrm e^{\mathrm i 2\Phi(\hat {\mathbf n})}\sum_k f(\nu_k)A(\hat {\mathbf n},\hat {\mathbf n}_A)\mathrm e^{\mathrm i 2\pi\mathbf u_k\cdot \hat {\mathbf n}}\right]\].
Discrete Fourier Transform, which is
the process of $\mathcal O(\mathcal N^2)$, can be carried out in 
$\mathcal O(\mathcal N \log_2 \mathcal N)$ with Fast Fourier Transform \citep{Numerical_Recipe_C,FastSphericalHarmonics}.
Since it is easiest to carry out Fast Fourier Transform on data of number which is a power of two, $\Delta \phi$ was chosen to be  $360^{\circ}/(2^{\lceil \log_2(2l_1)\rceil})$, where $\lceil\,\rceil$ denotes the smallest integer larger than or equal to the argument.
Choosing the optimal size for $\Delta \theta$ and $\Delta \phi$ and using Fast Fourier Transform enables the numerical computation of $b_{\pm2,lm}$ feasible in a reasonable amount of time even for interferometers of high $u$. 
We set the value for the integration cell to be zero and skips computing the rest of terms, when the beam function $A(\hat {\mathbf n},\hat {\mathbf n}_A)$ for the integration cell is smaller than $.1\%$ of its peak value. 
In integration over bandwidth, 
$\sum_k f(\nu_k)A(\hat {\mathbf n},\hat {\mathbf n}_A) e^{i 2\pi\mathbf B\frac{\nu_k}{c} \cdot \hat {\mathbf n}}$, an interference term of index $k$ is computed from a term of index $k-1$ as follows so that we can avoid computing time-consuming trigonometric function for each index $k$:
\begin{eqnarray*}
\lefteqn{\mathrm e^{\mathrm i 2\pi\frac{\nu_k}{c}\mathbf B\cdot \hat {\mathbf n}}=}\\
&&(\cos(\frac{2\pi\:\Delta \nu}{c} \mathbf B \cdot \hat{\mathbf n})+\mathrm i\sin(\frac{2\pi\:\Delta \nu}{c} \mathbf B \cdot \hat{\mathbf n}))
\mathrm e^{\mathrm i 2\pi\frac{\nu_{k-1}}{c} \mathbf B\cdot \hat {\mathbf n}},\end{eqnarray*}
where the computed value of $\cos(\frac{2\pi\:\Delta \nu}{c} \mathbf B \cdot \hat{\mathbf n})$ and $\sin(\frac{2\pi\:\Delta \nu}{c} \mathbf B \cdot \hat{\mathbf n})$
are repeatedly used. 
\section{Estimation in the absence of noise}
\label{no_noise}
\begin{figure}
\centering
\centering\includegraphics[scale=.5]{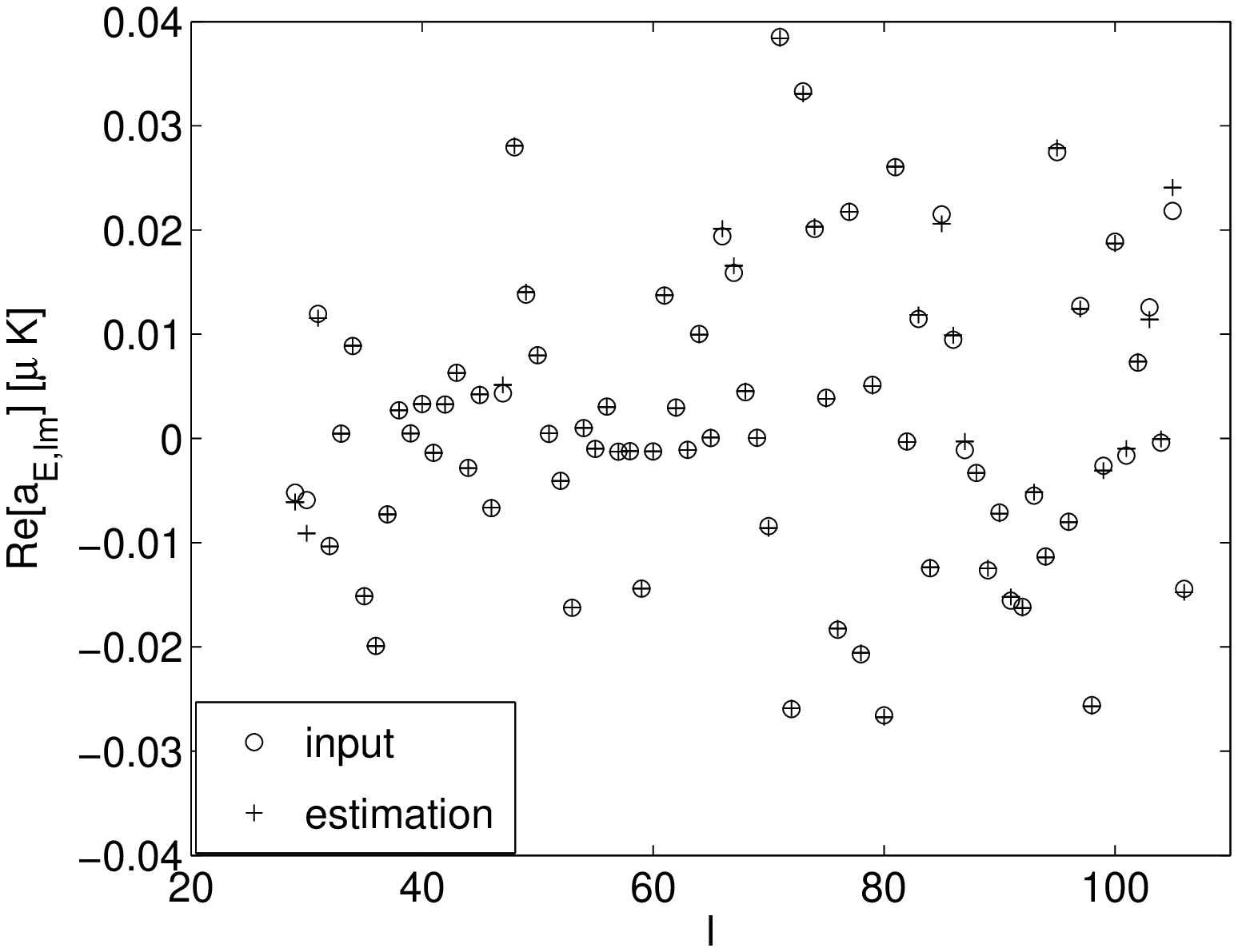}
\caption{estimated $\mathrm {Re}[a_{E,lm}]$ (m=29) in absence of noise}
\label{E_r-no_noise}
\end{figure}
\begin{figure}
\centering
\centering\includegraphics[scale=.5]{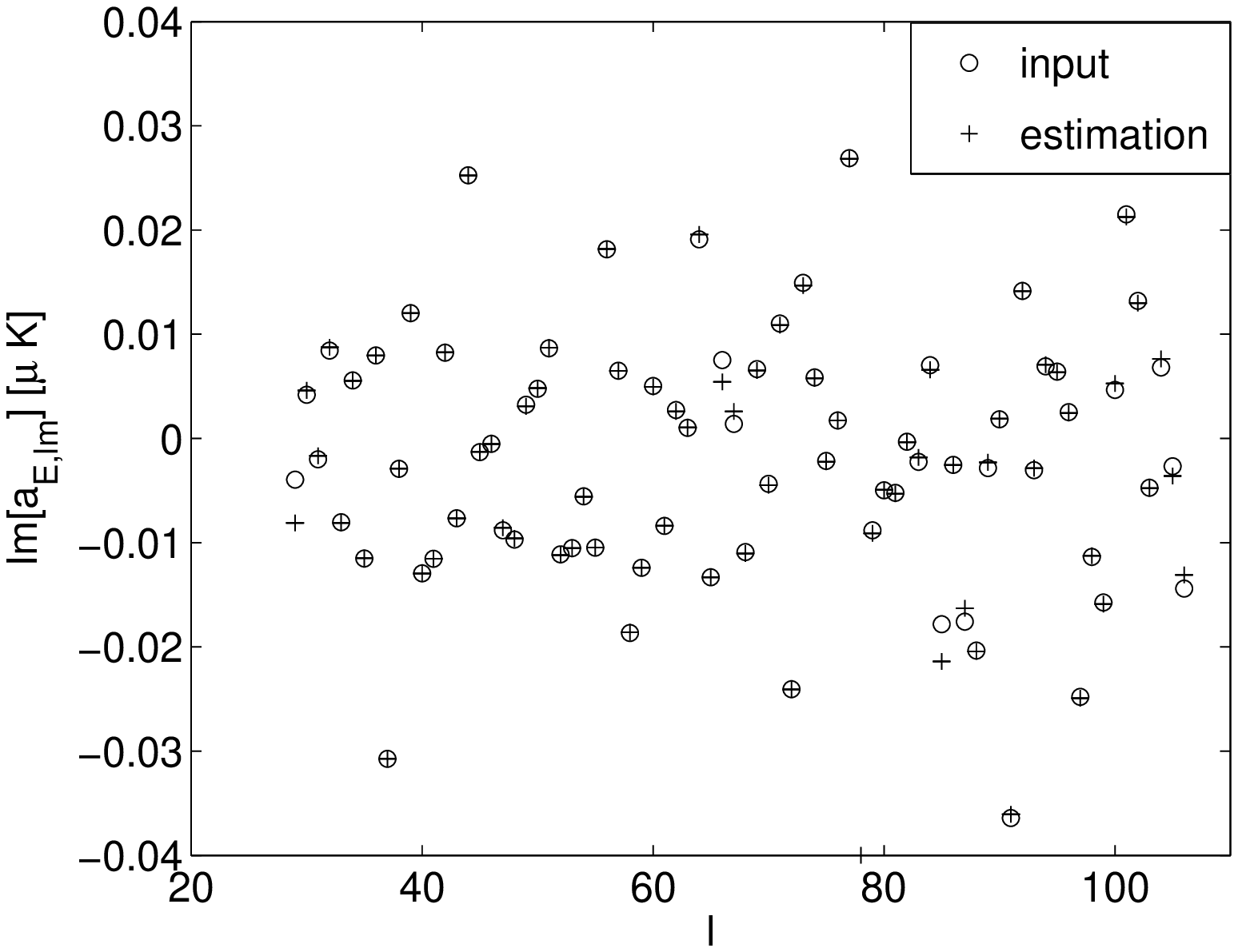}
\caption{estimated $\mathrm {Im}[a_{E,lm}]$ (m=29) in absence of noise}
\label{E_i-no_noise}
\end{figure}
\begin{figure}
\centering
\centering\includegraphics[scale=.5]{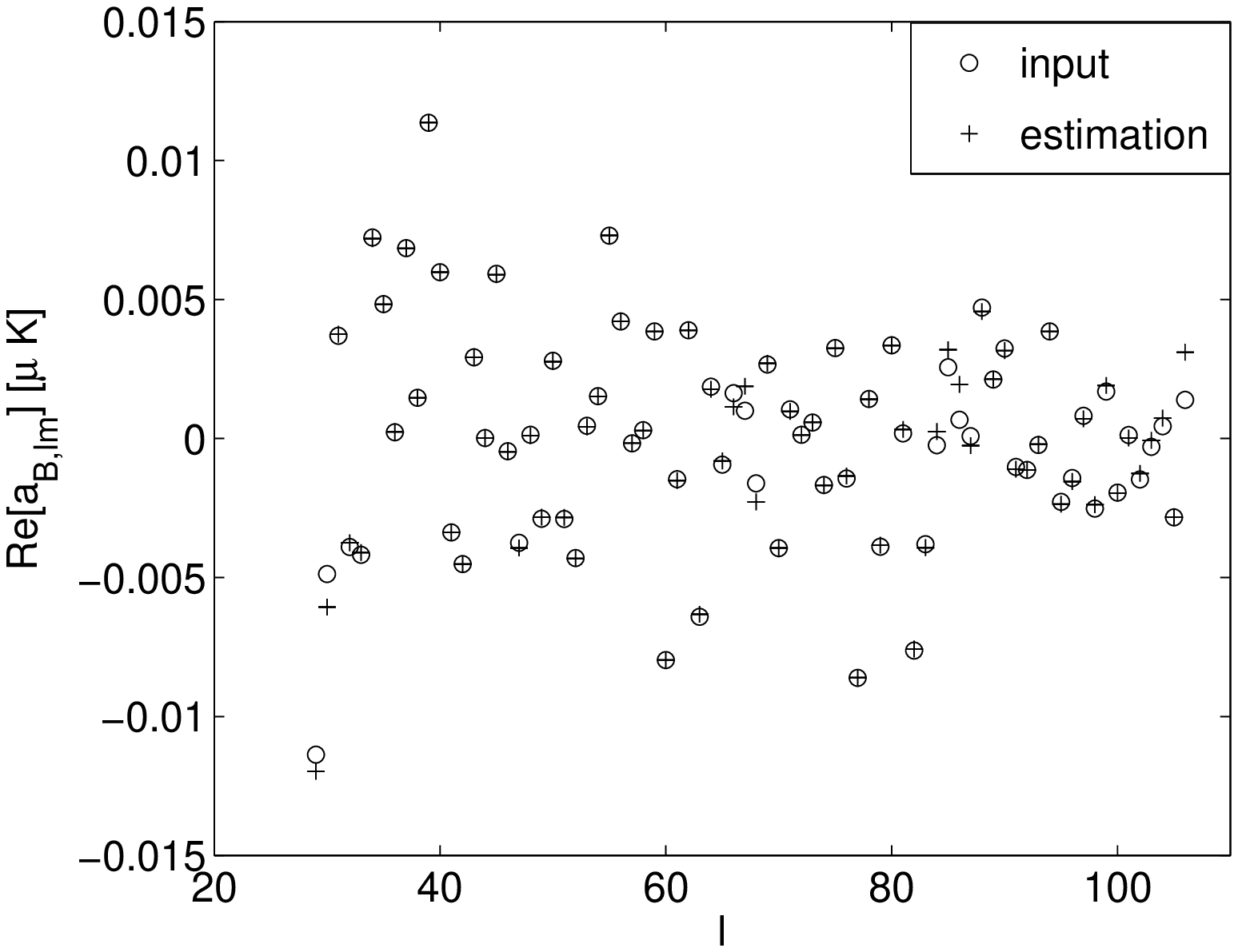}
\caption{estimated $\mathrm {Re}[a_{B,lm}]$ (m=29) in absence of noise}
\label{B_r-no_noise}
\end{figure}
\begin{figure}
\centering
\centering\includegraphics[scale=.5]{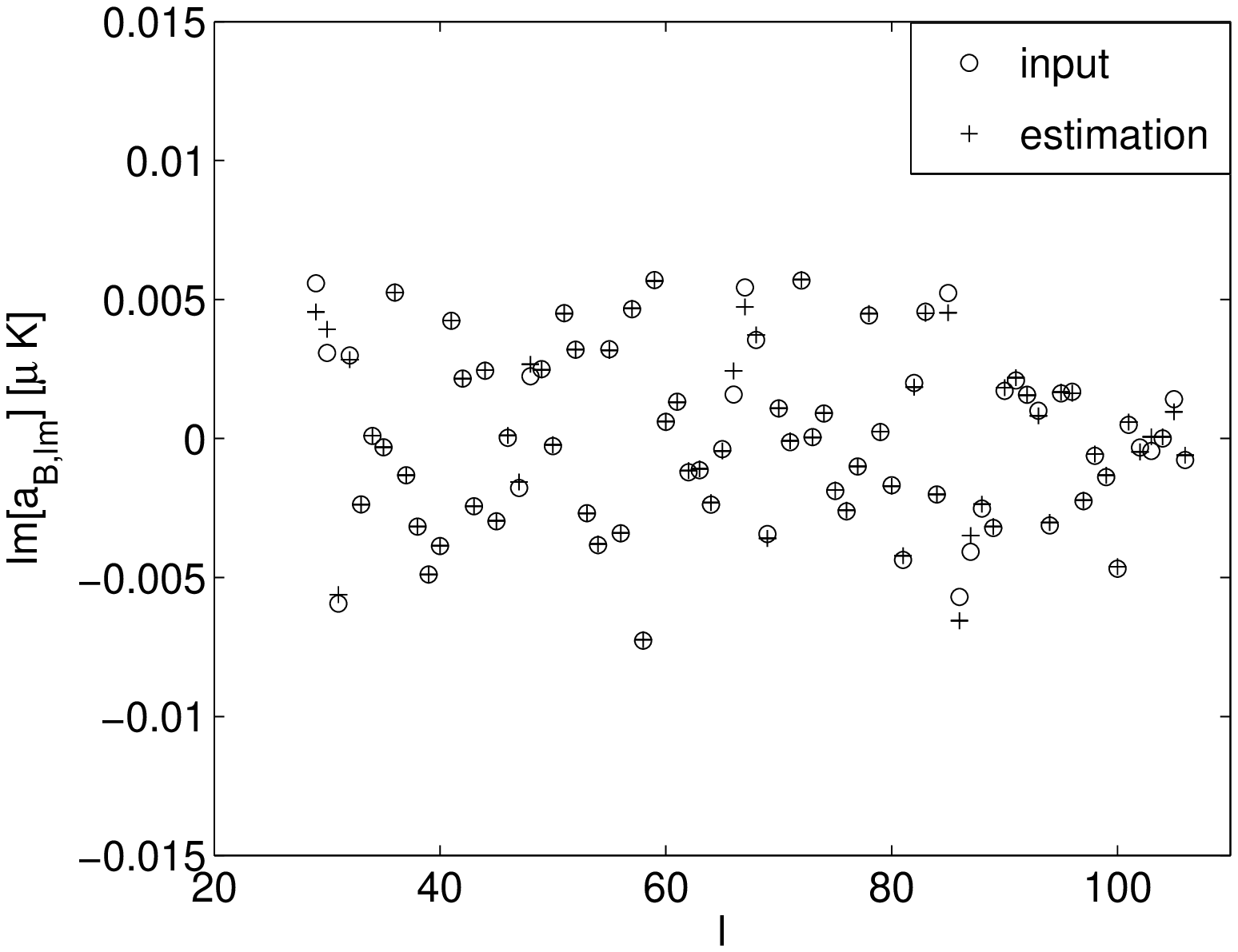}
\caption{estimated $\mathrm {Im}[a_{B,lm}]$ (m=29) in absence of noise}
\label{B_i-no_noise}
\end{figure}

Estimated $a_{2,lm}$ ($29 \le l \le 106,\;\;m=29,\,0,\,-29)$ in absence of noise
are shown together with the input $a_{2,lm}$ value in Fig. \ref{E_r-no_noise}, \ref{E_i-no_noise}, \ref{B_r-no_noise}, and \ref{B_i-no_noise}.
We assumed the same configuration with the simulated observation in \S \ref{simulation} except for the absence of noise. Small discrepancies between estimation and the input values, in spite of no noise, are attributed to residual error. As mentioned in \S \ref{simulation}, the residual error results from the contribution of $a_{2,lm}$ in the multipoles outside the cutoff region, since we determined $a_{2,lm}$ 
only over the multipoles where the window function is greater than 1\% of its peak value.
Some features of methods to facilitate $b_{2,lm}$ computation, which are discussed in Appendix \ref{computing_b2lm}, sacrifice the numerical precision, which also contributes to the discrepancies.

\label{lastpage}
\end{document}